\documentclass[12pt]{article}

\usepackage{graphicx}
\usepackage{algorithm,algorithmicx,algpseudocode}
\usepackage{enumerate}
\usepackage{amsfonts}
\usepackage{latexsym,amsmath}

\usepackage{slashbox}
\usepackage{amssymb,array}
\begin{document}
\title{\large \bf Cryptanalysis of Sun and Cao's Remote Authentication Scheme with User Anonymity}

\author{\small Dheerendra Mishra\thanks{E-mail:~{dheerendra@maths.iitkgp.ernet.in} }\\
\small Department of Mathematics,\\
\small Indian Institute of Technology Kharagpur,\\
\small  Kharagpur 721302, India\\}

\date{}
 \maketitle

\begin{abstract}
Dynamic ID-based remote user authentication schemes ensure efficient and anonymous mutual authentication between entities. In 2013, Khan et al. proposed an improved dynamic ID-based authentication scheme to overcome the security flaws of Wang et al.'s authentication scheme. Recently, Sun and Cao showed that Khan et al. does not satisfies the claim of the user's privacy. Moreover, They proposed an efficient authentication scheme with user anonymity. The Sun and Cao's scheme achieve  improvement over Khan et al.'s scheme in both privacy and performance point of view. Unfortunately, we identify that Sun and Cao's scheme does not resist password guessing attack. Additionally, Sun and Cao's scheme does not achieve forward secrecy.


\end{abstract}
\textbf{keywords:} {Smart card; Remote user authentication; Anonymity.}

\section{Introduction}{\label{intro}}
Advancement in technology has provided a scalable platform for various online services. In these services, a user access the remote servers via a public channel. However, an adversary is considered powerful enough that he can intercept, modify, delete and replace the messages, which transmits via public channel. This increase the threats to data security and integrity. One of the countermeasure is remote user authentication protocols~\cite{chien2002efficient,fan2005robust,jaspher2012smart,khan2013more, kumari2013cryptanalysis,li2013enhanced}. Most of the existing remote user authentication schemes ensure mutual authentication and session key agreement where the user and the server can mutually authenticate each other and draw a session key.

To achieve efficient and secure remote user services, low-cost smart card based authentication protocol has been presented to achieve a scalable solution. Most of the existing smart card based schemes face various kinds of attacks such as insider attack, password guessing attack, man-in-the middle attack, impersonation attack, repay attack and many more. Moreover, user anonymity is usually  not protected~\cite{boyd2003protocols,madhusudhan2012dynamic}. In 2009, Wang et al.~\cite{wang2009more} presented an efficient and secure dynamic ID-based remote user authentication scheme to achieve authorized and anonymous communication. However, in 2011, Khan et al.~\cite{khan2011cryptanalysis} identified the flaws in Wang et al.'s scheme and proposed and improvement to remove these flaws. In recently published paper, Sun and Cao's demonstrated that Khan et al.'s dynamic ID-based remote user authentication scheme claim to provide user anonymity is failing as an adversary can track any legitimate user's identity by eavesdropping on the mutual authentication session over the public channel. Additionally, they proposed an improved scheme and claimed that their scheme is reasonable in privacy, security and performance aspect.

In this article, we analyze the security of Sun and Cao's smart card based remote user authentication scheme. We identify that Sun and Cao's scheme is vulnerable to the password guessing attack. Moreover, we present the inefficiency of Sun and Cao's scheme to ensure forward secrecy.




The rest of the paper is organized as follows: Section \ref{notations} explains the meaning of symbols that are used in this paper. The brief review of Sun and Cao's scheme is given in Section \ref{review}. Section \ref{crypt} demonstrates the flaws in Sun and Cao's scheme. Finally, the conclusion is drawn  in Section \ref{conclusion}.

\section{Notations}\label{notations}


\begin{table}[H]
  \caption{Meaning of symbols of used notations that used throughout the paper}\label{t1}
\begin{tabular}{l|l}
\hline
Notation & Descryption\\
\hline
$U_i$ & User $i$\\
$S$ & A trustworthy server\\

$ID_i$ & Unique identity of user $i$\\
$PW_i$  & Unique password of user $i$\\
$N$     & Registration time of user\\
$T_U$ & Timestamp generated by $U_i$\\
$T_S$ & Timestamp generated by $S$\\
$sk$ & Session key\\
$x$  & Master key of $S$\\

$h(\cdot)$ \& $h_1(\cdot)$ &  One-way hash functions\\

$\oplus$ & XOR\\
$||$ & String concatenation operation\\

 \hline
\end{tabular}
\end{table}

\renewcommand{\labelitemi}{$ $}
\section{Review of Sun and Cao's Scheme}\label{review}
In 2013, Sun and Cao~\cite{sun2013privacy}  proposed an improvement of Khan et al.'s scheme~\cite{khan2011cryptanalysis} dynamic ID-based remote user authentication scheme. This comprises following phases:
\begin{itemize}
  \item[a.] Registration phase
  \item[b.] Login Phase
  \item[c.] Authentication Phase
  \item[d.] Lost smart card revocation Phase
\end{itemize}

In this section, we will briefly discuss registration, Login and authentication  phases of Sun and Cao's scheme as we only analyzes these phases in our study.
\renewcommand{\labelitemi}{$\bullet $}
\subsection{Registration Phase}
A new user can register with the server and gets the personalized smart card as follows:

\begin{description}

\item { Step 1.} $U_i$ chooses his identity $ID_i$ and password $PW_i$, and generates a random number $r$. Then, he computes $RPW = h(r||PW_i)$ and sends the message $<ID_i, RPW>$ to $S$ via secure channel.

\item[\bf Step 2.] Upon receiving the registration request, $S$ verifies the credentials of $ID_i$ and checks whether $ID_i$ exist or not in its database. If $ID_i$ exist and registers with any other user, it asks new identity. Otherwise, if $U_i$ does not exist in the server's database, $S$ computes $J = h(x||ID_i||N)$ and $L = J\oplus RPW$, where $N$ is the number of times $U$ registered. Then, $S$ embeds $\{L\}$ into smart card and secretly issues the smart card to $S$.

\item [\bf Step 3.] Upon receiving the smart card, $U_i$ stores $r$ into the smart card. %

\end{description}

\subsection{Login Phase}
When a user wants to login to the server, he inserts his smart card into the card reader and inputs the identity $ID_i$ and password $PW_i$, then login phase works  as follows:

\begin{description}

\item[\bf Step 1.] The smart card computes $RPW = h(r||PW_i)$ and $J = L\oplus RPW$.

\item[\bf Step 2.] The smart card computes $C_1 = h_1(J||T_U)$. Then $U_i$ sends $M_1 = <C_1, T_U>$ to $S$.

\end{description}


\subsection{Authentication Phase}
When server $S$ receives the login message, then server verifies the authenticity of the user. The user also verifies the server authenticity as follows:

\begin{description}

\item [\bf Step 1.] Upon receiving $U$'s message $M_1$, $S$ verifies the validity of timestamp $T_U$. If $T_U$ is incorrect, it terminates the session. Otherwise, $S$ searches
$ID_i$ among stored identities by verifying the condition $C_1 =? ~ h_1(h(x||ID_i||N)||T_U)$. If $ID_i$ verification does not hold for any identity of the database, S terminates the login session. Otherwise, S accepts the login request.

\item[\bf Step 2.] $S$ computes $C_2 = h_1(h(x||ID_i||N)||C_1||T_S)$. Then, $S$ sends the message $M_2 = <C_2, T_S>$ to $U_i$. Moreover, $S$ computes the session key $sk = h_1(J||C_2)$.

\item[\bf Step 3.] Upon receiving the message  $M_2$, $U_i$ verifies the timestamp $T_S$. If the verification does not hold, it ends the session. Otherwise, it verifies $C_2 =?~ h_1(J||C_1||T_S)$. If verification does not succeed, the session terminated. Otherwise, $S$ is authenticated and session key $sk = h_1(J||C_2)$ is computed.

\end{description}



\renewcommand{\labelitemi}{$-$}
\section{Security Weaknesses of Sun and Cao's Scheme}\label{crypt}

\renewcommand{\labelitemi}{$\bullet$}
\subsection{Off-line password guessing attack}
The password guessing attack is the one of the most common attack on password based authentication protocols using smart card. An adversary can perform password guessing attack on  Sun and Cao's scheme as follows:  
%
%
%

 \begin{description}
 \item {\bf Step 1.} Guess the password $PW_i^*$, and compute $RPW^* = h(PW_i^*||r)$ and $J^* = L\oplus RPW^*$.

 \item {\bf Step 2.} Verify the condition $C_1 =? ~ h_1(J^*||T_U)$.

 \item {\bf  Step 3.} If the verification holds, the guessing of passwords is succeeding. Otherwise, repeat {\bf Step 1} and {\bf  Step 2}.
  \end{description}

\subsection{Forward secrecy}
In Sun and Cao's scheme, the session key $sk$ is the hashed output of user's long term secret key $J$ along with $C_2$, $i.e.$ $sk = h_1(J||C_2)$. This shows that Sun and Cao's scheme doe not ensure forward secrecy, as a compromise of user's long term secret key $J$ causes compromise of all the established session keys. This is possible as follows:

\begin{itemize}
  \item An adversary can achieve all the previously transmitted message via public channel, $i.e$, an adversary can achieve $C_2$.

  \item The adversary can compute the session key $sk = h_1(J||C_2)$.

\end{itemize}

\section{Conclusion}\label{conclusion}
An efficient password based remote user authentication scheme should resist all kinds of attack. In this article, we have demonstrated that Sun and Cao's privacy preserving remote user authentication scheme is vulnerable to off-line password guessing attack and fails to ensure forward secrecy.



\end{document}